\documentclass[aps,prl,twocolumn,superscriptaddress,10pt,amssymb]{revtex4-2} 
\usepackage{microtype}
\usepackage[varbb,varg,smallerops,upint]{newtx} 
\usepackage{bm,dsfont}

\AtBeginDocument{\renewcommand{\vec}[1]{\bm{#1}}}

\usepackage[dvipsnames,svgnames]{xcolor}
\usepackage{hyperref}
\hypersetup{colorlinks=true, linkcolor=NavyBlue, urlcolor=NavyBlue, citecolor=NavyBlue} 

\usepackage{physics,braket,slashed} 
\usepackage[capitalise]{cleveref}

\usepackage{graphicx}
\graphicspath{{figures/}}

\usepackage{tabularray}
\UseTblrLibrary{varwidth}

\begin{document}


\title{Exciton-based sensing of remote electron correlations in 2D heterostructures}

\author{Tobias\ M.\ R.\ Wolf}
\affiliation{Department of Physics, University of Texas, Austin, Texas, 78712, USA}
\author{Tian Xie}
\affiliation{Department of Physics, University of California at Santa Barbara, Santa Barbara, CA, 93116, USA}
\author{Chenhao Jin}
\affiliation{Department of Physics, University of California at Santa Barbara, Santa Barbara, CA, 93116, USA}
\author{Allan\ H.\ MacDonald}
\affiliation{Department of Physics, University of Texas, Austin, Texas, 78712, USA}

\date{\today} 

\begin{abstract}
Many monolayer transition metal dichalcogenides, including MoS$_2$, MoSe$_2$, WS$_2$, and WSe$_2$, are direct bandgap two-dimensional (2D) semiconductors with sharp optical 
resonances at excitonic bound state frequencies. 
Recent experiments have demonstrated that excitonic
resonance frequencies in multilayer van der Waals stacks are altered by long-range Coulomb interactions with electrons in nearby but electrically isolated 2D materials. 
These modulations have been successfully used to detect transitions between
distinct states of remote strongly correlated 2D electron fluids. 
In this Letter we provide a theory of these frequency shifts, enabling a more 
quantitative interpretation of excitonic-sensing experiments, and apply it as an
example to WSe$_2$ that is proximate to graphene bilayers and multilayers.
\end{abstract}

\maketitle


\paragraph{Introduction ---}
Monolayer transition-metal dichalcogenides (TMDs) host tightly bound excitons whose optical 
resonances dominate \cite{mak2016photonics,mueller2018exciton,regan2022emerging} 
visible and near-infrared optical absorption, reflectance, and luminescence spectra.  
Because of the long-range nature of Coulomb interactions between electrons, 
exciton energies of two-dimensional semiconductors vary strongly with 
the surrounding three-dimensional environment, sometimes shifting by hundreds of 
meV when substrates, overlayers, or spacers are changed 
\cite{raja2017coulomb,hsu2019dielectric}.  This extreme sensitivity to remote layers has 
been harnessed to probe many-body phenomena in proximate electron systems of interest.
For example, Popert \textit{et al.}\ established exciton spectroscopy as a contact-free 
probe of electronic correlations by demonstrating that fractional-quantum-Hall states in 
graphene layers manifest as abrupt shifts in the exciton lines of neighboring WSe$_2$ 
layers~\cite{popert2022optical}. 

\begin{figure} 
\centering
\includegraphics{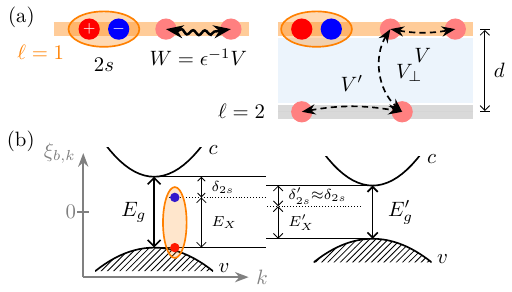}
\caption{Optical sensing of proximate layers using 
semiconductor bandgap renormalization. 
(a)~Side view of an exciton (ellipse) in semiconductor layer $\ell=1$, in the absence 
(left) and presence (right) of electrons in a proximate $\ell=2$ layer separated by 
distance $d$. The arrows indicate electronic intra- and interlayer interactions. 
(b)~Schematic band structure and exciton energies for each case. 
For weakly bound excitons, the change in binding energy ($\delta_{2s}$) is negligible. The resonance shift ($\delta E_X$) then measures the bandgap renormalization ($\delta E_g$).
}
\label{fig1}
\end{figure}

Figure~\ref{fig1} illustrates the underlying mechanism of excitonic sensing.
Charge response in a remote
layer screens the Coulomb interaction between electrons and holes in the TMD.
This typically decreases both the bandgap of the semiconductor layer, and the binding energies of the states in the exciton Rydberg series.  
For the tightly bound $1s$ exciton these two effects nearly cancel, but for the
weakly bound excitons used in sensing experiments the binding energy is reduced by an 
order of magnitude and the band-gap shift dominates \footnote{When the change in binding energy is not negligible for the most weakly bound visible exciton, the bandgap can be estimated by extrapolating from the exciton Rydberg series.}.
Exploiting this principle, Xie \textit{et al.}\ recently used $2s$ exciton reflection contrast measurements in monolayer WSe$_2$ as a 
contactless probe of spin–valley polarization transitions and integer quantum-Hall ferromagnetism in adjacent graphene bilayers and multilayers \cite{Xie2025NatCommun}.
Notably, there is a mysterious even/odd filling-factor alteration in the 2s exciton energy 
as the bilayer graphene's $N=0$ Landau-level octet is filled that lacks an explanation.

The goal of this Letter is to provide a theory that allows exciton 
spectroscopy to serve as a \emph{quantitative} probe; what exactly are the 
exciton energy shifts measuring about the remote layer?  
We find that the answer to this 
question is remarkably simple when the characteristic length scales of the system of 
interest are long compared to the host crystal lattice constants. 
First-principles multilayer formalisms are 
successful in capturing exciton energy shifts due to 
screening by insulating substrates and encapsulants \cite{NaikJain_PRB_2017,NaikJain_PRM_2018,Winther2017_GdW,thygesen2017calculating}.
Our goal is to account for the additional corrections that arise when
a remote layer hosts a \emph{strongly correlated} electron fluid, and the 
many-body state of that fluid changes.
We derive a compact transparent expression that
links the band-gap shifts in a TMD semiconductor layer to the \emph{full
interacting dynamic charge susceptibility} $\chi(\vec q,\omega)$ of the
electron fluid in the remote layer.  As examples of the utility of
this formula, we evaluate energy shifts at spin--valley ferromagnetic phase transitions
in rhombohedral multilayer graphene states and show that they reproduce experimental 
observations, and explain the even/odd variation in the $N=0$ Landau level of bilayer graphene.  Finally, we discuss other potential applications of exciton sensing.

\begin{figure}
\centering
\includegraphics[width=\linewidth]{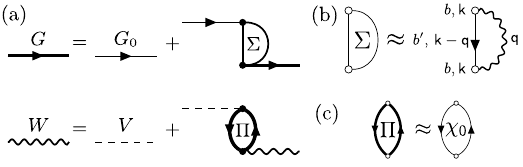}
\caption{
Feynman diagrams for the $G_0W$ approximation. 
(a) Dyson equations for the full Green's function $G$ with self-energy $\Sigma$, and the dressed interaction $W$ with irreducible polarizability $\Pi$. 
(b) Diagonal self-energy $\Sigma_b$ for band $b$ and 
$\mathsf{k}=(\vec{k},i\omega_n)$, where $\vec{k}$ is the momentum and $\omega_n$ is a (fermionic) Matsubara frequency, in the $G_0W$ approximation.
We implicitly integrate/sum over internal variables, such as $\mathsf{q}=(\vec{q},i\Omega_m)$, where $\Omega_m$ is a (bosonic) Matsubara frequency.  
(c) In the $G_0W_0$ approximation the irreducible polarizability is replaced by the free-particle susceptibility $\chi_0$.
}
\label{fig:feynman}
\end{figure}

\paragraph{Theory of the proximity-induced bandgap shift ---}
We build our theory on the well-established $G_0W_0$ method, which has been applied 
with great success to two-dimensional (2D) semiconductors \cite{RevModPhys.74.601,PhysRevB.87.235132,PhysRevB.34.5390,thygesen2017calculating,G.D.Mahan1990,giuliani2008quantum}. 
The key simplifying observation that allows for analytic progress is the
observation that instantaneous changes in charge density in the remote layer with
2D wavevector $\vec{q}$ yield a potential in the semiconductor layer that is suppressed
by a factor of $\exp(-|\vec{q}|d)$; atomic scale charge density changes in the
remote layer are therefore invisible in the probe layer and vice versa.  Since the 
remote layer or layers are separated vertically from the probe layer
by many 2D lattice constants, only coarse-grained changes 
in electron density averaged over unit cells are relevant.
Since potentials that vary slowly on an atomic scale do not couple different bands,
the change in self-energy due to the presence of the remote band is accurately
diagonal in band and for a given band will depend only on electron propagators 
with the same band index.  To calculate the change in the bandgap we need 
only to evaluate the difference between the real electron self-energy changes
at conduction and valence band extrema:
\begin{equation}
\label{eq:gwse} 
\Delta\Sigma_{b\vec{k}}(i\omega_n) 
= 
-\frac{1}{\beta}\sum_{m} \underset{\scriptscriptstyle\mathrm{BZ}}{\int} \, \!\frac{d^2 q}{(2 \pi)^2}  G^{(0)}_{b\vec{k}-\vec{q}}(i \omega_n-i \Omega_m) 
    \; \Delta W_{\vec{q}}(i \Omega_m),
\end{equation}
where $\beta$ is the inverse temperature, $\omega_n$ ($\Omega_m$) are fermionic (bosonic) Matsubara frequencies, and ${m\in\mathbb{Z}}$. 
In \cref{eq:gwse}, $\Delta W_{\vec{q}}(i \Omega_m)$ is the frequency and 
wave-vector dependent spatially coarse-grained change in the dynamically screened potential,
$  G^{(0)}_{b\vec{k}}(i \omega_n)   = (i\omega_n -\xi_{b\vec{k}})^{-1} ,$ 
is the non-interacting Matsubara Green's function, and 
$\xi_{b\vec{k}}$ is the band energy relative to the chemical potential $\mu$.
When a remote layer is present, an electron in the semiconductor layer 
interacts directly with other electrons in that layer and also with
the charge densities induced in the remote layer by its response to the first semiconductor layer electron
\footnote{Since $W$ depends in a coordinate representation on the positions of the 
two interacting particles in the semiconductor layer, its Fourier representation a matrix in momentum space for every $\vec{q}$ in the Brillouin-zone.
If atomic scale spatial variations were relevant in $\Delta W$, 
the fact that the semiconductor layer and the remote layer have different lattices would complicate its description. $\Delta W$ would have the 
lattice periodicity of the remote crystal and this would 
complicate its matrix elements in the representation of semiconductor layer Bloch states.  Fortunately, the $\exp(-|\vec{q}+\vec{G}|d)$ factors in the interactions between layers make these complications irrelevant.}.  It follows that 
\begin{align}\label{eq:Vbarecorrection}
\Delta W_{\vec{q}}(i\Omega_m) = \chi_{\vec{q}}^{\mathrm{cf}}(i\Omega_m)
\, (V_{\vec{q}}^{\perp})^2.
\end{align}
In \cref{eq:Vbarecorrection} $V_{\vec{q}}^{\perp}$ is
the interlayer Coulomb interaction for 2D 
electrons in an anisotropic dielectric environment:
$V_{\vec{q}}^{\perp} =  2\pi e^2/(\epsilon_{\text{eff}} |\vec{q}|) \exp(- |\vec{q}|\kappa d)
$
with $\epsilon_{\text{eff}}=(\epsilon_{\parallel}\epsilon_{zz})^{1/2}$ and $\kappa=(\epsilon_{\parallel}/\epsilon_{zz})^{1/2}$
\footnote{When the TMD is directly adjacent to the  correlated fluid layer, we use the dielectric constants of the TMD (in case of WSe$_2$, $\epsilon_{\parallel}\simeq 15$ and $\epsilon_{zz}\simeq 8$), and in presence of a hBN spacer, we use $\epsilon_{\parallel}\simeq 7$ and $\epsilon_{zz}\simeq 3$.}. 
Combining \cref{eq:gwse,eq:Vbarecorrection} with a spectral representation 
for $\Delta W$ allows the Matsubara summation to be performed (see Supplemental Material), and gives the following simple expression for the band extremum on-shell self-energies:
\begin{widetext}
\begin{align}\label{eq:gwse_ret}
\Delta\Sigma^{R}_{b\vec{k}}
=
-
\int \frac{d^{2}q}{(2\pi)^{2}}
\!
\int_{0}^{\infty}\!\frac{d\omega}{\pi}
\Biggl[ \;
  \mathrm{Im}\Delta W_{\vec{q}}(\omega+ i0^+) \;
  \frac{\bar n_{b\vec{k}+\vec{q}} + n_{B}(\omega)}
       {\Delta\xi^{b\vec{k}+\vec{q}}_{b\vec{k}} - \omega + i0^+} \; 
+  \;
   \mathrm{Im}\Delta W_{-\vec{q}}(\omega+ i0^+) \; 
  \frac{n_{b\vec{k}+\vec{q}} + n_{B}(\omega)}
       {\Delta\xi^{b\vec{k}+\vec{q}}_{b\vec{k}} + \omega + i0^+} \; 
\Biggr].
\end{align}
\end{widetext}
In \cref{eq:gwse_ret} $\Delta\xi^{b'\vec{k}+\vec{q}}_{b\vec{k}}=\xi_{b\vec{k}} - \xi_{b'\vec{k}+\vec{q}}$, $n_{b\vec{k}}=n_F(\xi_{b\vec{k}})$ with Fermi occupation factor $n_F(\xi)$, and ${\bar{n}=1-n}$.  We have used that $\langle \psi_{b\vec{k}-\vec{q}}|e^{-i\vec{q}\cdot\vec{\hat{r}}}|\psi_{b\vec{k}}\rangle \approx 1$ for $\vec{q}$ small compared to the BZ size.  Since we are interested in the low-temperature limit ($T\to 0$), we set the Bose distribution function [$n_B(\omega)$] to zero from here forward.

\begin{figure*}
\centering
\includegraphics[width=1.0\linewidth]{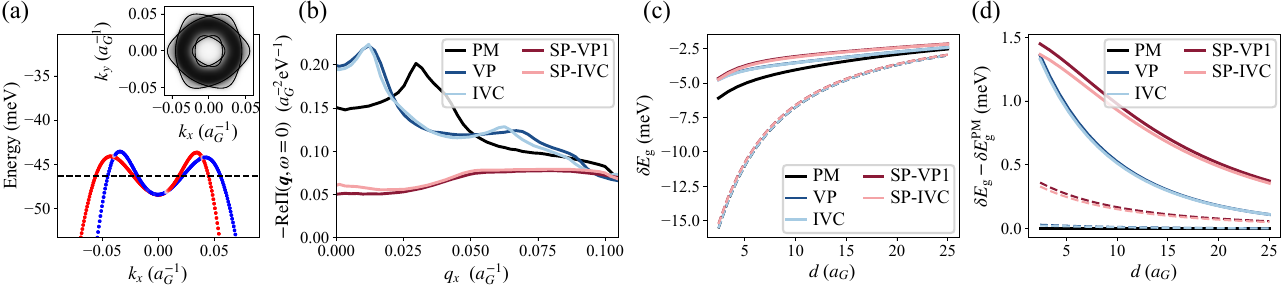}
\caption{
Rhombohedral trilayer graphene (RTG) at large displacement field $U_D = 30$~meV and density $n_e = -0.97\times10^{12}$~cm$^{-2}$, and the induced bandgap shift in proximate WSe$_2$. 
(a) Mean-field electronic bands of the symmetric (paramagnetic; PM) state (red: valley $K$, blue: valley $K'$) and Fermi surface (inset). 
(b) Real part of the momentum-dependent static RPA polarizability along $q_x$ for competing flavor orders—PM, spin-polarized with valley polarization (SP-VP), spin-polarized with inter-valley coherence (SP-IVC), valley-polarized (VP), and inter-valley coherent (IVC). 
(c) Induced WSe$_2$ bandgap shift versus interlayer spacing $d$ for the same set of metastable Hartree–Fock flavor states; values are absolute (i.e., not referenced to the PM state). 
Dashed curves show the static approximation [cf.~\cref{eq:finalenergygap3}], and solid curves include the retardation correction [cf.~\cref{eq:finalenergygap_corr}]. 
(d) WSe$_2$ bandgap change relative to PM, i.e. $\delta E_g-\delta E_g^{\mathrm{PM}}$.
We used a six-band-per-flavor continuum model, $300\times300\times3000$ momentum–frequency grids, and dual-gate screening (see SM for details), and $a_G$ is the graphene lattice constant.
}
\label{fig:trilayer_graphene_0field}
\end{figure*}

For the sake of definiteness, we now specialize to
the case of a semiconductor layer with a direct gap at the 
triangular-lattice BZ corners $\vec{K}$. 
In the limit $T\to 0$, all valence-band states are occupied, and all conduction-band states are empty. From \cref{eq:gwse_ret}, we see that the conduction band shift $\Delta\Sigma_{c\vec{K}}$ is negative, and the valence band shift $\Delta\Sigma_{v\vec{K}}$ is positive, so that the bandgap is reduced: 
\begin{align}
\label{eq:energygap}
&\delta E_{\text{g}} 
= \Re\Delta\Sigma_{c\vec{K}} - \Re\Delta\Sigma_{v\vec{K}}
\\
&= -\int
     \!\!\frac{d^2 q}{(2 \pi)^2} 
     \int_{0}^{\infty} \! d \omega \, 
\left[ 
  \frac{(V_{\vec{q}}^{\perp})^2 \,\Delta_{\vec{q}}(\omega)}{\xi_{c\vec{K}-\vec{q}}-\xi_{c\vec{K}}+\omega}
+
  \frac{(V_{\vec{q}}^{\perp})^2 \,\Delta_{-\vec{q}}(\omega)}{\xi_{v\vec{K}}-\xi_{v\vec{K}-\vec{q}}+\omega}
\right] \! ,
\nonumber 
\end{align}
where $\Delta_{\vec{q}}=-\pi^{-1} \mathrm{Im}\chi^{\text{cf}}_{\vec{q}}$ is a non-negative spectral function. 
\Cref{eq:energygap} is our main result. 
When the band dispersion in the semiconductor layer is weak compared to that of the 
correlated fluid layer the energy differences between the band states at wavevectors $\vec{K}$ and $\vec{K}-\vec{q}$ can be neglected [$\Delta \Sigma_{b\vec{k}}^{(b)}\approx 0$ in \cref{eq:deltagwseB}]. In this instantaneously-screened limit, the frequency
integral can be performed analytically using the spectral representation of 
$\Delta W_{\vec{q}}(\omega= 0^{+})$ and the bandgap shift expression simplifies to: 
\begin{align}
\label{eq:finalenergygap3}
    \delta E_{\text{g}}^{\mathrm{stat}} 
&= 
\int \!\!
 \frac{d^2 q}{(2 \pi)^2} \, \Delta W_{\vec{q}}(\omega= 0^{+}). 
\end{align}
This static shift can be identified as the change in the exchange self-energy of states 
near the valence band edge relative to those at the conduction band edge.
The difference between \cref{eq:finalenergygap3} and \cref{eq:energygap} can
be viewed as a correction due to the retarded character of the remote layer response: 
\begin{align}\label{eq:finalenergygap_corr}
\delta E_{\text{g}}^{\mathrm{dyn}}
&= \int
     \!\!\frac{d^2 q}{(2 \pi)^2} 
     (V_{\vec{q}}^{\perp})^2
     \int_{0}^{\infty} \! d \omega \, 
     \frac{\Delta_{\vec{q}}(\omega)}{\omega}
\nonumber 
\\
& \quad
\times
\bigg[ 
  \frac{\xi_{c\vec{K}-\vec{q}}-\xi_{c\vec{K}}}{\xi_{c\vec{K}-\vec{q}}-\xi_{c\vec{K}}+\omega}
+
\frac{\xi_{v\vec{K}}-\xi_{v\vec{K}+\vec{q}}}{\xi_{v\vec{K}}-\xi_{v\vec{K}+\vec{q}}+\omega}
\bigg]. 
\end{align}
We evaluate these contributions separately because the frequency integral in \cref{eq:finalenergygap_corr}, which converges more rapidly at large $\omega$
than that in \cref{eq:energygap}, is more convenient for numerical evaluation.
 
\emph{Rhombohedral Multilayer Graphene ---} 
As a first example of optical sensing, consider the case of a monolayer WSe$_2$ semiconductor layer with effective mass ${m_v\approx m_c \approx 0.3 m_e}$, proximate to 
strongly correlated metallic rhombohedrally-stacked multilayer graphene.
The latter is known to undergo transitions to half and quarter metal 
phases with spontaneous spin--valley flavor ($\alpha\in\{K\uparrow,K\downarrow,K'\uparrow,K'\downarrow\}$) polarizations
upon sweeping electronic density $n_e$ at large transverse displacement field $D$
\cite{Weitz2010,Velasco2012,Bao2011,Lee2014,Chen2019Mott,Chen2019SC,Chen2020Chern,Yang2022Spectroscopy,Shi2020RG,Zhou2021HalfQuarter,Zhou2021HalfQuarter,zhang_spinorbit_2022,Zhou2022Isospin,lu2023fractional,Han2024Correlated,Han2023Multiferroic,Han2024QAH,wolf2024magnetismdilute2deg,Zhang2010ABC,Koshino2009Trigonal,Huang2023Magnetism,Qin2023FRG,Cea2022KohnLuttinger,Dong2023Isospin,Ghazaryan2021Annular,Koh2024SOC,Lu2022RG,Dong2024SpinCanting}. 
As recently observed \cite{Xie2025NatCommun}, flavor polarization transitions
are accompanied by significant bandgap shifts, see  
\cref{fig:trilayer_graphene_0field}.  
To evaluate \cref{eq:energygap}, we first express the 
charge susceptibility $\chi^{\text{cf}}$ of the proximate graphene multilayer in terms of its polarizability $\Pi^{\text{cf}}$: 
\begin{align}\label{eq:chi_RPA}
    \chi^{\text{cf}}_{\vec{q}}(\omega) = \frac{\Pi^{\text{cf}}_{\vec{q}}(\omega)}{1-V_{\vec{q}}\Pi^{\text{cf}}_{\vec{q}}(\omega)},
\end{align}
where $ V_{\vec{q}} =  (2\pi e^2)/(\sqrt{\epsilon_{\parallel}\epsilon_{zz}} |\vec{q}|)
$ is the intralayer Coulomb interaction. 
We evaluate $\Pi^{\text{cf}}(\vec{q},\omega)$ using the random-phase approximation (RPA) 
and realistic continuum models with trigonal warping and 
particle-hole asymmetry.  ($N$-layer rhombohedral graphene has 
$2N$ $\pi$-bands per flavor -- see SM for model details.) 
\footnote{We use a phenomenological dielectric constant $\epsilon_r = 35$ in
the Hartree-Fock calculation to obtain a valley polarization phase diagram that is approximately consistent.  Otherwise we use realistic dielectric functions.
We also account for gate-screening by using interaction $V_{\text{eff}}(\vec{q})=2\pi e^2 \tanh(d_{\text{gate}} q)/(\epsilon_r q)$ with $d_{\text{gate}}=50~\textrm{nm}$}.
Here, we consider displacement field $U_D=30$~meV at density $n_e=-0.97\times 10^{12}\,\mathrm{cm}^{-2}$, see Figs.~\ref{fig:trilayer_graphene_0field}(a). 
Near this density, the Hartree-Fock equations have multiple competing metastable solutions: paramagnetic, and half-metal and quarter-metal states, see \cref{fig:trilayer_graphene_fp}. 
The symmetric state has an annular Fermi surface in each valley, and the real part of the polarizability has abrupt features due to Fermi surface nesting, see \cref{fig:trilayer_graphene_0field}(b). The imaginary part of the dynamic polarizability is shown in \cref{fig:trilayer_graphene_fp,fig:rtg-chi0-three-panel}, and has strong contributions at small frequencies and finite momenta, which is related to the phase space for particle-hole pairs across the Fermi surface. The resulting RPA susceptibility features plasmons.
We consider the case where the sensing layer and RTG are separated by an interlayer distance $d$ and again evaluate \cref{eq:finalenergygap3,eq:finalenergygap_corr}: in Figs.~\ref{fig:trilayer_graphene_0field}(c) 
and \cref{tab:bandshift_rtg}, we show the sensing layer's bandgap change induced by a flavor transition from the symmetric state into each symmetric broken candidate state in graphene. We again observe that retardation corrections from the dynamic susceptibility substantially reduce the bandgap change naively expected from static screening. Figures~\ref{fig:trilayer_graphene_0field}(c,d) 
also illustrates how the factor $\exp(-qd)$ in $V_\perp$ suppresses the bandgap change when interlayer separation reaches the length scale set by the Fermi surface in RTG. 

\begin{figure}
\centering
\includegraphics[width=1.0\linewidth]{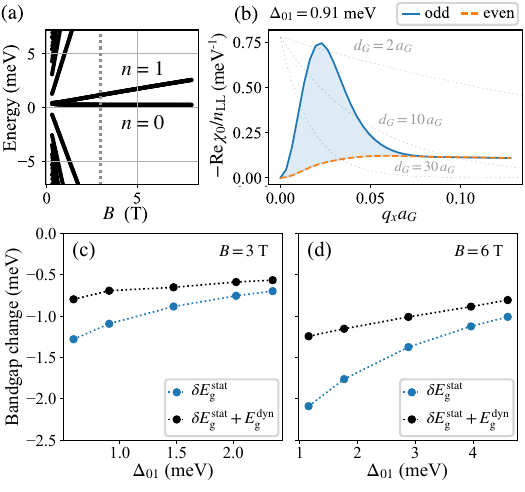}
\caption{%
Landau levels and susceptibility of bernal bilayer graphene, and the resulting even–odd bandgap shift of a proximate monolayer WSe$_2$. 
(a)~Landau‐level (LL) spectrum versus magnetic field $B$. 
(b)~Static susceptibility $\mathrm{Re}\,\chi_0$ at $B=3$~T versus $q_x$ at $q_y=0$ for fillings $\nu=0,1,2,3,4$; shading highlights the extra ${0\!\to\!1}$ contribution present at odd filling. Gray dotted lines indicate how the finite interlayer spacing $d$ suppresses contributions to the sensing‐layer band shift. The corresponding dynamic susceptibility is shown in \cref{fig:bilayer_graphene_chi0_filling0,fig:bilayer_graphene_chi0_filling1}. 
(c–d)~Even–odd bandgap shift in monolayer WSe$_2$ when the BBG filling changes from even $(\nu=0)$ to odd $(\nu=1)$, plotted versus the LL separation $\Delta_{01}$, for (c) $B=3$~T and (d) $B=6$~T. Blue lines: static‐screening contribution only; black lines: static+dynamic. 
Unless stated otherwise we use a Landau‐level cutoff $N_{\rm cut}=70$; see SM for details.
}

\label{fig:bilayer_graphene_LL_suscept}
\end{figure}

\emph{Bernal Bilayers in a finite magnetic field ---} 
The Landau level (LL) spectrum of Bernal bilayer graphene has an octet of low-energy states, formed by Landau orbitals $n=0$ and $n=1$ for each spin-valley flavor, see \cref{fig:bilayer_graphene_LL_suscept}(a). The octet degeneracy is lifted by trigonal warping and exchange interactions~\cite{KousaMacDonald2025_blg} that both influence the splitting $\Delta_{01}$ between $n=0$ and $n=1$ states, and can furthermore lead to spontaneous spin--valley polarization 
\cite{Zhao2010PRL,Weitz2010,Maher2013NatPhys,Lee2014,Hunt2017NatCommun,Lee2016NanoLett,Yin2019PRL,Lee2013NanoLett,Datta2017NatCommun,Xie2025NatCommun,Zhang2012PRB,Zhang2011PRL,Kharitonov2012PRL,KousaMacDonald2025_blg}. Here, we treat $\Delta_{01}$ as phenomenological parameter which we control with the trigonal warping parameter $\gamma_4$ in our bandstructure model (see SM). 
The splitting $\Delta_{01}$ has a large effect on the polarizability of BBG in each spin--valley flavor: at odd filling fractions --- when the chemical potential is in the gap between ${n=0}$ and ${n=1}$ levels --- additional particle-hole excitations near momenta $q\sim \ell_B^{-1}$ contribute $\sim n_{\mathrm{LL}}/\Delta_{01}$ to the static RPA polarizability, see \cref{fig:bilayer_graphene_LL_suscept}(b), where $\ell_B$ is the magnetic length and $n_{\mathrm{LL}}=(2\pi\ell_B^2)^{-1}$ the Landau level degeneracy. These flavor-diagonal particle-hole excitations contribute strongly to the dynamic RPA susceptibility, as illustrated in \cref{fig:bilayer_graphene_chi0_filling0,fig:bilayer_graphene_chi0_filling1}, where they lead to a low-frequency peak $\sim \delta(\omega-\Delta_{01}) n_{\mathrm{LL}}$ near $q\sim \ell_B^{-1}$. 

We use the RPA polarizability per flavor $\alpha$ in \cref{fig:bilayer_graphene_LL_suscept} together with \cref{eq:chi_RPA,eq:finalenergygap3,eq:finalenergygap_corr} to calculate
the bandgap differences between even and odd LL filling factors within the octet.  
Assuming a spin-valley polarized state at integer fillings ${\nu=-1}$ and ${\nu=1}$, we approximate the flavor-summed RPA polarizability as ${\Pi^{\mathrm{cf},\nu=-1}\approx 4\,\Pi_{\alpha}^{\mathrm{cf},\nu_{\alpha}=0}}$ and ${\Pi^{\mathrm{cf},\nu=0}\approx 3\,\Pi_{\alpha}^{\mathrm{cf},\nu_{\alpha}=0}+\Pi_{\alpha}^{\mathrm{cf},\nu_{\alpha}=1}}$, respectively. For example, at ${B=3~\text{T}}$ with ${\Delta_{01}=0.91~\text{meV}}$, we find that the bandgap in the sensing layer changes by $\left.\Delta E_g(B=3~\textrm{T})\right|_{\nu=-1}^{\nu=1}=-0.7~\textrm{meV}$ at odd filling factors
because of their partial orbital occupations.

In \cref{fig:bilayer_graphene_LL_suscept}(c,d), we show the dependence of the bandgap shifts in \cref{eq:chi_RPA,eq:finalenergygap3,eq:finalenergygap_corr} at two magnetic field strengths $B=3~\text{T}$ and $B=6~\text{T}$ as a function of LL splitting $\Delta_{01}$. We find that while neglecting the dynamic screening contribution qualitatively captures the experimentally observed even--odd effect \cite{Xie2025NatCommun}, it overestimates the band shifts in the sensing layer (blue lines). Including dynamic screening (black lines) substantially reduces the overall bandgap change, and leads to more moderate dependence on the LL splitting $\Delta_{01}$. 

\emph{Discussion ---} In this Letter we presented a 
theory of proximity-induced bandgap shifts in a 2D semiconductor caused by a correlated electron fluid in an adjacent layer.  Our key result is a simple expression for the semiconductor conduction- and valence-band on-shell self-energy corrections in the presence of an external correlated fluid.  We have evaluated these spectator layer energy shifts
numerically for two examples of correlated fluid physics of recent interest - the cases of Bernal bilayer graphene at finite magnetic fields and 
rhombohedral trilayer graphene at large displacement fields. 
A central finding is that the retardation correction to the screening can substantially reduce the bandgap shift below its static approximation value, especially when the correlated electron fluid has prominent low-frequency collective modes or particle-hole excitations 
with momenta smaller than the inverse interlayer distance. 

Our theory can be extended in several useful directions.  
First, accounting for the finite thickness of graphene
multilayer stacks is straightforward and would allow our theory to describe optical signatures
in the sensing layer related to changes in the layer polarization of the 
probed multilayers.  Our treatment can also be extended to account for 
proximity-induced spin-orbit coupling in graphene layers adjacent to TMD layers. 
Both effects become relevant at small interlayer distances. 
Second, we focused on the renormalization of the semiconductor bandgap, building on experimental evidence that the $2$s exciton binding energy is much less sensitive to the dielectric environment than the $1$s exciton. Including \cref{eq:Vbarecorrection} as a correction to the Wannier-Mott equation could allow to account for this effect. 
In future work, the present framework may be applied to other many-body phenomena
of fundamental interest in strongly correlated 2D materials, 
including flavor ferromagnetism, fractional quantum Hall (FQH) physics and superconductivity.
In FQH states the full interacting dynamic structure factor, is a much-studied quantity.
Our work shows that it is directly probed by excitonic sensing.
Finally, exotic superconducting states are common in 2D 
materials and can~\cite{kee1998polarizability} also lead to  
strong changes in dynamic density-response functions.  Our work shows that 
combined transport and optical sensing studies have the potential to shed new light on
these surprising states which are difficult to probe.

\begin{acknowledgments}
\emph{Acknowledgments ---} A.H.M. and T.M.W. were supported by the U.S. Department of Energy Office of Science, Office of Basic Energy Sciences under award DE-SC0021984.
\end{acknowledgments}

\bibliography{references}


\clearpage
\onecolumngrid

\setcounter{secnumdepth}{2}
\setcounter{equation}{0} 
\setcounter{figure}{0} 
\setcounter{table}{0} 
\setcounter{page}{1}
\renewcommand\thepage{SM-\arabic{page}}
\makeatletter
\renewcommand{\theequation}{S\arabic{equation}}
\renewcommand{\thefigure}{S\arabic{figure}}
\renewcommand{\thetable}{S\arabic{table}}
\renewcommand{\bibnumfmt}[1]{[S##1]}

\setlength{\parindent}{0pt} 
\setlength{\parskip}{6pt plus 2pt minus 1pt} 

\section*{Supplementary Material}

\subsection{On-shell retarded self-energy}
\label{sec:onshell_SE}

Within the one–shot $G_{0}W$ approximation the change of the electronic
self-energy produced by inserting a remote screening layer is, on the
imaginary-frequency axis,
\begin{equation}
\label{eq:S1}
\Delta\Sigma_{b\vec{k}}\!\left(i\omega_{n}\right)
=
-\frac{1}{\beta}
\sum_{m}
\underset{\scriptscriptstyle\mathrm{BZ}}{\int}
\!\!\frac{d^{2}q}{(2\pi)^{2}}
\sum_{b'\vec{G}\vec{G}'}
\rho^{b'\vec{k}-\vec{q}}_{\,b\vec{k},\vec{G}}
\rho^{b'\vec{k}-\vec{q}\,*}_{\,b\vec{k},\vec{G}'}
\,
G^{(0)}_{b'\vec{k}-\vec{q}}\!\bigl(i\omega_{n}-i\Omega_{m}\bigr)
\,
[\Delta W_{\vec{q}}\!\bigl(i\Omega_{m}\bigr)]_{\vec{G}\vec{G}'},
\end{equation}
where $b$ labels a band, $\vec{k}$ a Bloch wave-vector,
$\omega_{n}$ ($\Omega_{m}$) are fermionic (bosonic) Matsubara
frequencies, and
$\beta\!=\!1/k_{\mathrm{B}}T$.
The non-interacting Green function is
$G^{(0)}_{b\vec{k}}(i\omega_{n})
  =(i\omega_{n}-\xi_{b\vec{k}})^{-1}$, and the Bloch-wave density
form factor 
$
\rho^{b'\vec{k}-\vec{q}}_{\,b\vec{k},\vec{G}}
=
\bigl\langle u_{b'\vec{k}-\vec{q}}\bigl|e^{i\vec{G} \cdot \hat{\vec{r}}}\bigr|u_{b\vec{k}}\bigr\rangle
$ 
at reciprocal vector $\vec{G}$. 
The screened interaction in coordinate representation is $\Delta W(\vec{r}_1,\vec{r}_2) = \int \frac{d^dq}{(2\pi)^d}\sum_{\vec{G},\vec{G}'} e^{i(\vec{q} + \vec{G})\cdot\vec{r}_1}\, e^{-\,i(\vec{q} + \vec{G}')\cdot\vec{r}_2} \, [\Delta W_{\vec{q}}]_{\vec{G},\vec{G}'}$, using standard convention for Fourier transform for functions lattice-periodic in two variables, i.e., $f(\vec{r}_1+\vec{a},\vec{r}_2+\vec{a})=f(\vec{r}_1,\vec{r}_2)$.

\paragraph{Spectral representation and Matsubara sum.}
The screened interaction is written in spectral form
\begin{align}
\label{eq:S3}
[\Delta W_{\vec{q}}(i\Omega_{m})]_{\vec{G}\vec{G}'}
=
-\int_{0}^{\infty}\!\!\frac{d\omega}{\pi}\,
\Bigl[
  \frac{[\Delta W''_{\vec{q}}(\omega)]_{\vec{G}\vec{G}'}}{i\Omega_{m}-\omega}
 -\frac{[\Delta W''_{-\vec{q}}(\omega)]_{\vec{G}'\vec{G}}}{i\Omega_{m}+\omega}
\Bigr],
&& 
[\Delta W''_{\vec{q}}(\omega)]_{\vec{G}\vec{G}'}
\equiv
\Im\!\bigl[\Delta W_{\vec{q}}(\omega+i0^{+})\bigr]_{\vec{G}\vec{G}'},
\end{align}
and we can introduce a structure factor $\Delta W''_{\vec{q}}(\omega)=A^{-1} (1-e^{-\beta\omega}) S_{\vec{q}}(\omega)$, where $A$ is the system area. We note that $[\Delta W_{\vec{q}}(-\omega)]_{\vec{G}\vec{G}'}=[\Delta W_{\vec{q}}(\omega)]_{\vec{G}'\vec{G}}^*$ for real frequencies $\omega$.

Inserting the spectral representation \cref{eq:S3} and using the identity \cite{giuliani2008quantum}
\begin{align*}
\frac{1}{\beta} \sum_{m=-\infty}^{\infty} \frac{1}{i \omega_n-i \Omega_m-\xi} \frac{1}{i \Omega_m-\omega}
=-\frac{n_F\left(-\xi\right)+n_B\left(\omega\right)}{i \omega_n-\xi-\omega}
=\frac{n_F\left(\xi\right)+n_B\left(-\omega\right)}{i \omega_n-\xi-\omega},
\end{align*}
we perform the Matsubara summation in Eq.~\eqref{eq:S1}.  Here, $n_F(\xi)$ [$n_B(\omega)$] are Fermi-Dirac [Bose-Einstein] functions, and they have some useful relations: $n_F(\xi)+n_F(-\xi)=1$ and $n_B(\omega)+n_B(-\omega)=-1$, and $n_F(-\xi)-n_F(\xi)=1-2 n_F(\xi)=\tanh(\beta \xi/2)$ and $n_B(\omega)-n_B(-\omega)=1+2 n_B(\omega)=1/\tanh(\beta \omega/2)$. We will occasionally use the short-hand notation $n_{\mu}=n_F(\xi_{\mu})$.

After 
analytic continuation ($i\omega_{n}\!\to\!\xi_{b\vec{k}}+i\eta$) the
\emph{retarded} on-shell shift becomes \cref{eq:gwse_ret} of the main text, i.e., 
\begin{align}\label{eq:S4}
\Delta\Sigma^{R}_{b\vec{k}}
&=
-
\underset{\scriptscriptstyle\mathrm{BZ}}{\int}
\!\!\frac{d^{d}q}{(2\pi)^{d}}
\!
\sum_{b'\vec{G},\vec{G}'}
\rho^{b'\vec{k}-\vec{q}}_{b\vec{k},\, \vec{G}}
\rho^{b'\vec{k}-\vec{q}\;*}_{b\vec{k},\, \vec{G}'}
\!
\int_{0}^{\infty}\!\frac{d\omega}{\pi}
\Biggl[
  [\Delta W''_{\vec{q}}(\omega)]_{\vec{G}\vec{G'}}\,
  \frac{\bar n_{b'\vec{k}-\vec{q}} + n_{B}(\omega)}
       {\Delta\xi^{b'\vec{k}-\vec{q}}_{b\vec{k}} - \omega + i\eta}
+ 
  [\Delta W''_{-\vec{q}}(\omega)]_{\vec{G}'\vec{G}}\,
  \frac{n_{b'\vec{k}-\vec{q}} + n_{B}(\omega)}
       {\Delta\xi^{b'\vec{k}-\vec{q}}_{b\vec{k}} + \omega + i\eta}
\Biggr],
\end{align}
where $\Delta\xi^{b'\vec{k}-\vec{q}}_{b\vec{k}}=\xi_{b\vec{k}}-\xi_{b'\vec{k}-\vec{q}}$. 
When the semiconductor
and the correlated-fluid layer are separated by many lattice constants
one may safely neglect local-field effects by keeping only
$\vec{G}=\vec{G}'=\vec{0}$ in Eq.~\eqref{eq:S4}. As discussed in the main text, the screening contribution to the interaction is related to the susceptibility through $\Delta W_{\vec{q}}=(\epsilon_{\vec{q}}^{-1}-\mathds{1}) V_{\vec{q}}=V_{\vec{q}}\chi_{\vec{q}}V_{\vec{q}}$, such that $\Delta W_{\vec{q}}''=V_{\vec{q}}\chi_{\vec{q}}''V_{\vec{q}}$.

There are several ways to decompose \cref{eq:S4} into separate contributions. In addition to the dynamic/static split discussed in the main text, one may also separate into a screened exchange contribution, 
\begin{subequations} \label{eq:deltagwse3} 
\begin{align} \label{eq:deltagwseA} 
\Delta \Sigma_{b\vec{k}}^{(a)}
= 
& \int
     \!\!\frac{d^2 q}{(2 \pi)^2} 
\sum_{b'}\sum_{\vec{G}\vec{G}'}
\rho^{b'\vec{k}-\vec{q}}_{\,b\vec{k},\vec{G}}
\rho^{b'\vec{k}-\vec{q}\,*}_{\,b\vec{k},\vec{G}'}
\, \left(\frac{1}{2} - n_F(\xi_{b'\vec{k}-\vec{q}})\right)
\Delta W_{\vec{q}}(\xi_{b\vec{k}}-\xi_{b'\vec{k}-\vec{q}}+i\eta),
\end{align}
where we used \cref{eq:S3} for the frequency integral in \cref{eq:gwse_ret}, and a second contribution related to bosonic fluctuations/plasmons 
\begin{align} \label{eq:deltagwseB} 
&\Delta \Sigma_{b\vec{k}}^{(b)}
= 
- \int
     \!\!\frac{d^2 q}{(2 \pi)^2} 
     \sum_{b'}\sum_{\vec{G}\vec{G}'}
\rho^{b'\vec{k}-\vec{q}}_{\,b\vec{k},\vec{G}}
\rho^{b'\vec{k}-\vec{q}\,*}_{\,b\vec{k},\vec{G}'}
     \int_0^{\infty} \!\frac{d\omega}{\pi}  
     \left(\frac{1}{2} + n_B(\omega)\right)
\left[ 
  \frac{[\Delta W_{\vec{q}}(\omega)]_{\vec{G}\vec{G}'}}{\xi_{b\vec{k}}-\xi_{b'\vec{k}-\vec{q}}-\omega+i\eta}
+
  \frac{[\Delta W_{-\vec{q}}(\omega)]_{\vec{G}'\vec{G}}}{\xi_{b\vec{k}}-\xi_{b'\vec{k}-\vec{q}}+\omega+i\eta} 
\right].
\end{align}
\end{subequations}

\subsection{Electrostatic potential of a finite-width dielectric slab}

Consider a slab extending from $z=-\tfrac{w}{2}$ to $z=+\tfrac{w}{2}$ with anisotropic permittivities $\bigl(\epsilon_{\parallel},\epsilon_{\perp}\bigr)$.  
The exterior ($\lvert z\rvert>w/2$) has $\bigl(\epsilon'_{\parallel},\epsilon'_{\perp}\bigr)$.  
A uniform line charge of density $\lambda$ fills the slab in the $z$-direction (i.e.\ $\rho(x,y,z)=\lambda\,\delta(x)\,\delta(y)\,\Theta\bigl(z+\tfrac{w}{2}\bigr)\,\Theta\bigl(\tfrac{w}{2}-z\bigr)$), normalized to $\lambda=1/w$.  
In Gaussian units, Gauss's law is $\nabla\cdot\bigl(\vec{\epsilon}\,\nabla\Phi\bigr)=-4\pi\,\rho.$

After a 2D Fourier transform (wavevector $q$ in-plane) the general solution becomes
\begin{equation*}
\Phi(z) 
= 
\begin{cases}
A\,e^{\,q\,\kappa\,z} \;+\; B\,e^{-\,q\,\kappa\,z} \;+\; c, &  \lvert z\rvert<\tfrac{w}{2},\\[6pt]
F\,\exp\!\bigl[-\,q\,\kappa'\,\bigl(z - \tfrac{w}{2}\bigr)\bigr], & z>\tfrac{w}{2},\\[6pt]
E\,\exp\!\bigl[\,q\,\kappa'\,\bigl(z + \tfrac{w}{2}\bigr)\bigr], & z<-\tfrac{w}{2},
\end{cases}
\quad
\text{with}
\quad
c \;\equiv\; \frac{4\pi\,\lambda}{\epsilon_{\parallel}\,q^2} \;=\; \frac{2\pi}{\epsilon_{\parallel} q}\,\frac{1}{q\bigl(w/2\bigr)},
\end{equation*}
and $\kappa=\sqrt{\epsilon_{\parallel}/\epsilon_{\perp}}$, $\epsilon = \epsilon_{\perp}\,\kappa$, and similarly for $\kappa'$ and $\epsilon'$.
Continuity of $\Phi$ and $\epsilon_{\perp}\,\frac{\partial\Phi}{\partial z}$ at $z=\pm\tfrac{w}{2}$ yields a linear system solved for $A,B,E,F$.  
Substituting these back gives the closed-form solution everywhere.  
For a symmetric slab with identical exterior media and $\kappa'$, we find
\begin{equation}
\Phi(z) 
= 
\begin{cases}
 \displaystyle
 \frac{2\pi}{\epsilon \, q}\,\frac{1}{\bigl(\tfrac{q\kappa w}{2}\bigr)}
\biggl[
1 
 - 
 \frac{\cosh(q\,\kappa\,z)}
      {\frac{\epsilon}{\epsilon'}\,\sinh(\tfrac{q\,\kappa\,w}{2})
       + \cosh(\tfrac{q\,\kappa\,w}{2})}
\biggr],
& \lvert z\rvert<\tfrac{w}{2},
\\[10pt]
 \displaystyle
 \frac{2\pi}{\epsilon' q}\,\frac{\tanh(\tfrac{q\,\kappa\,w}{2})}{\bigl(\tfrac{q \kappa w}{2}\bigr)}
 \frac{1}
      {1 + \frac{\epsilon}{\epsilon'}\,\tanh(\tfrac{q\,\kappa\,w}{2}) }
\,e^{-\,q\,\kappa'\,(|z|-\tfrac{w}{2})},
& |z| >\tfrac{w}{2}.
\end{cases}
\end{equation}
We see that in the limit $w\to 0$, the interlayer potential becomes $\Phi(z)\to 2\pi/(\epsilon'q) \exp(-q\kappa'|z|)$, as expected.

\subsection{Irreducible polarizability in RPA}

In a transverse magnetic field $B$, we can write the single-particle basis in terms of Landau level states as $\ket{\psi_{n X }}=\sum_{n',\tau} \psi_{nX;n'X\tau} \ket{\tau}\otimes\ket{n'X}$, where $\tau$ is an additional spin or orbital degree of freedom. The RPA polarizability then is 
\begin{align}\label{eq:polarizability_LL}
    \Pi^{\mathrm{B}}_{\vec{q}}(\omega) 
    &= \frac{1}{2\pi \ell^2} \sum_{n, m} \frac{n_n-n_m}{\xi_{n}-\xi_{m}+\omega+i \delta}
   \; |\rho^{nm}_{\vec{q}}|^2,
\qquad
\rho^{nm}_{\vec{q}} 
=
\braket{\psi_{n}|e^{-i\vec{q}\cdot\vec{r}}|\psi_{m}}
= \sum_{n'
    m'} \left(\sum_{\tau} \psi_{n;n'\tau}^* \psi_{m;m'\tau}\right) M^{n'm'}_{\vec{q}},
\end{align}
where $\ell$ is the magnetic length, $n,m$ labels the eigenvalues,  $\mathcal{M}^{nm}_{\vec{q}}$ is the momentum-transfer form factor, and $\xi_{n}$ are Landau level energies relative to the chemical potential $\mu$, and $\psi_{n}$ are Landau level wavefunctions. The form factor is related to  
\begin{align}
M^{nm}_{\vec{q}}
= \bra{n} e^{-i \vec{q}\cdot\vec{r}} \ket{m}
=
e^{i(\pi-\theta)\,|n-m|} \; 
\sqrt{\frac{\mathrm{min}(n,m)!}{\mathrm{max}(n,m)!}}
e^{-\frac{q^2\ell^2}{4}}
\left(\frac{q^2\ell^2}{2}\right)^{|n-m|/2}
L_{\mathrm{min}(n,m)}^{|n-m|}\left(\frac{q^2\ell^2}{2}\right),
\end{align}
where $\theta=\arg(q_x+i q_y)$, and $L^{\alpha}_n$ denotes generalized Laguerre polynomials. 

When the magnetic field is absent, we use the Bloch states $\ket{\psi_{n\vec{k}}}=\sum_{\vec{G},\tau} u_{n\vec{k};\tau\vec{G}} \ket{\tau}\otimes\ket{\vec{k}-\vec{G}}$, where $n$ is the band index, $\vec{k}$ is Bloch momentum, and $\ket{\vec{k}-\vec{G}}$ are plane-wave states, with band energies $\xi_{n \vec{k}}$. The RPA polarizability in this case is  
\begin{align}
\Pi_{\vec{q}}(\omega) 
&= \frac{1}{A} \sum_{n, m, \vec{k}} \frac{
n_{n \vec{k}} -
n_{m \vec{k}+\vec{q}} 
}{\xi_{n \vec{k}} - \xi_{m \vec{k}+\vec{q}} + \omega + i\delta} 
|\rho^{nm}_{\vec{k},\vec{k}+\vec{q}}|^2,
\quad
\rho^{nm}_{\vec{k},\vec{k}+\vec{q}} 
=
\braket{\psi_{n\vec{k}}|e^{-i\vec{q}\cdot\vec{r}}|\psi_{m\vec{k}+\vec{q}}}
=
\sum_{\vec{G},\tau} u^*_{n\vec{k};\tau\vec{G}} \, u_{m\vec{k}+\vec{q};\tau\vec{G}}.
\end{align}

In the main text, we calculated the wavefunctions $\psi_{nX;n'X\tau}$ and $u_{n\vec{k}}$ for multilayer graphene from numerical diagonalization as described in \cref{sec:rhomb_graphene_supmat}. 

To numerically evaluate $\Pi_{\vec{q}}(\omega)$, we closely follow the method outlined in Ref.~\cite{shishkin2006implementation}: we first compute the RPA spectral function 
\begin{align}
 \Im\Pi_{\vec{q}}(\omega) 
 = 
 -\frac{\pi \,\mathrm{sgn}(\omega)}{A} \sum_{n,m,\vec{k}} \delta(\omega-[\xi_{n \vec{k}}-\xi_{m\vec{k}+\vec{q}}])\, [n_{n \vec{k}}-n_{m\vec{k}+\vec{q}}]\, 
  |\!\braket{\psi_{n\vec{k}}|e^{-i\vec{q}\cdot\vec{r}}|\psi_{m\vec{k}+\vec{q}}}\!|^2,
\end{align}
using bin-search and interpolation, and then numerically evaluate the Hilbert transform 
\begin{align}
\Pi_{\vec{q}}(\omega+i \delta) 
= -\frac{1}{\pi}\int_{-\infty}^{\infty} \!\!\!\! d\omega' \, \frac{\Im\Pi_{\vec{q}}(\omega')}{\omega-\omega'+i \delta} 
= -\frac{1}{\pi}\int_{0}^{\infty} \!\!\!\! d\omega' \, 
\left[\frac{\Im\Pi_{\vec{q}}(\omega')}{\omega-\omega'+i \delta} 
- \frac{\Im\Pi_{-\vec{q}}(\omega')}{\omega+\omega'+i \delta} \right]
\, , 
\end{align}
to obtain the real part. To restrict ourselves to positive frequencies, we used $\Im\Pi_{\vec{q}}(-\omega)=-\Im\Pi_{-\vec{q}}(\omega)$ and note that generally $\Im\Pi_{-\vec{q}}(\omega)\neq \Im\Pi_{\vec{q}}(\omega)$, unless inversion symmetry is present.  We use a fine frequency resolution of $\delta\omega \sim 0.1$~meV over a frequency grid from $\omega=0$ to $\omega_{\text{max}}=\mathrm{max}(\xi_{m\vec{k}+\vec{q}}-\xi_{n\vec{k}})\sim 14$~eV.

\subsection{Continuum model for rhombohedral $N$-layer graphene}
\label{sec:rhomb_graphene_supmat}

\begin{table}[b]
\caption{\label{tab:graphene_params_ABC}Tight-binding parameters (in eV) for rhombohedral trilayer graphene, see also Refs.~\cite{Zhang2010ABC,Koshino2009Trigonal,Zhou2021HalfQuarter}.}
\centering
\begin{tblr}{
  width=0.8\linewidth,
  colspec = {*{8}{X}},
  rowsep = 1pt,
  column{1} = {leftsep=0pt},
  column{Z} = {rightsep=0pt},
  hline{1,Z} = {1pt,solid},
  hline{2,4,6} = solid,
}
 $\gamma_0$ & $\gamma_1$ & $\gamma_2$ & $\gamma_3$ & $\gamma_4$ & $U$ & $\Delta$ & $\delta$ \\
 $3.160$ & $0.380$ & $-0.015$ & $-0.290$ & $0.141$ & $0.030$ & $-0.0023$ & $-0.0105$ \\
\end{tblr}
\end{table}

\textit{Continuum model.} Rhombohedral $N$-layer graphene is a stack of $N$ graphene layers (each with sublattices $A$ and $B$), with in-plane lattice constant $a=2.46$~{\AA} and interlayer spacing $d=3.4$~{\AA}. Each successive graphene layer is laterally shifted so that \(A_j\) in layer \(j\) is directly above \(B_{j+1}\) in layer \(j+1\), with a stacking periodicity of 3 layers.
We use continuum model for low-energy $\pi$-electrons with spin $s=\pm 1/2$ near each nonequivalent valley $\tau K$ (with $\tau=\pm 1$), which leads to four `flavors` $\alpha=(s,\tau)$. Ordering the basis as $A_1,\,B_1,\,A_2,\,B_2,\dots,\,A_N,\,B_N$, the continuum Hamiltonian per flavor is 
\begin{align}\label{eq:continuum_hamiltonian_Nlayer}
    h_N(\vec{k}) = \begin{bmatrix}
    t(\vec{k})+U_1 & t_{12}(\vec{k}) & t_{13} & 0 & \cdots & 0\\
    t_{12}^\dagger(\vec{k}) & t(\vec{k})+U_2 & t_{12}(\vec{k}) & t_{13} & \ddots & \vdots\\
    t_{13}^\dagger        & t_{12}^\dagger(\vec{k}) & t(\vec{k})+U_3 & t_{12}(\vec{k}) & \ddots & 0\\
    0                     & t_{13}^\dagger        & t_{12}^\dagger(\vec{k}) & t(\vec{k})+U_4 & \ddots &  t_{13}\\
    \vdots                & \ddots                & \ddots                & \ddots                & \ddots & t_{12}(\vec{k})\\
    0                     & \cdots                     & 0                     &  t_{13}^\dagger                      & t_{12}^\dagger(\vec{k}) & t(\vec{k})+U_N
    \end{bmatrix}_{2N\times2N}.
\end{align}
The various hopping terms are defined compactly as
\begin{align}
    t(\vec{k}) &= \begin{bmatrix} 0 & v_0\,\pi^\dagger \\ v_0\,\pi & 0 \end{bmatrix}, \quad
    t_{12}(\vec{k}) = \begin{bmatrix} -v_4\,\pi^\dagger & v_3\,\pi \\ \gamma_1 & -v_4\,\pi^\dagger \end{bmatrix}, \quad
    t_{13} = \begin{bmatrix} 0 & \gamma_2/2 \\ 0 & 0 \end{bmatrix},
\end{align}
with $\pi=\tau k_x+ik_y$ and velocity parameters $v_i=(\sqrt{3}/2)a\gamma_i/\hbar$ ($i=0,3,4$), see \cref{tab:graphene_params_ABC}. The layer potentials $U_j$ (with $j=1,\dots,N$) account for the effect of external gates. Intrinsic Ising-type spin--orbit coupling (SOC) $\lambda$ is negligible (about $10$--$50$ $\mu$eV), but proximity to transition metal dichalcogenides (e.g., WSe$_2$ without spacer) can drastically enhance it in the nearest graphene layer (up to $\lambda \sim 0.800$ meV). In this work, we neglect SOC to focus on qualitative effects.

\begin{figure*}
    \centering
    \includegraphics[width=0.9\linewidth]{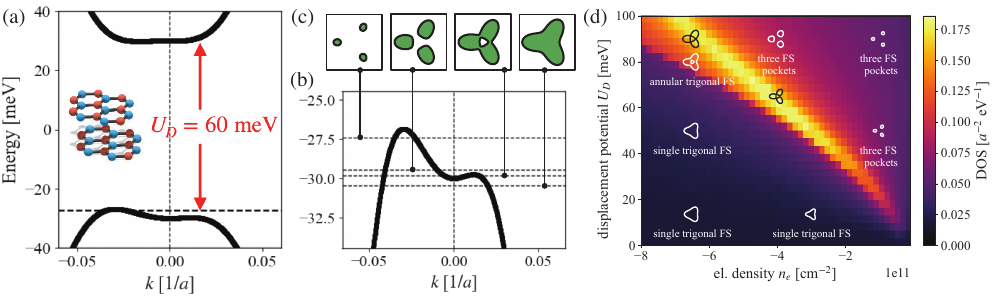}
    \caption{
    Non-interacting, trigonally-warped band structure of bernal bilayer graphene (sketched in inset) at large displacement fields. (a) Electronic bands near valley $K$ at displacement potential $U_D=60$ meV. (b) Same bands but zoomed in on the valence band to highlight the strong trigonal warping and (c) Fermi surfaces at different hole doping potential, illustrating Lifshitz transitions from pockets, to annular to single pocket. (b) Density of states (DOS) at the Fermi surface as function of electronic density $n_e$ and displacement potential $U_D$. Insets indicate the Fermi sea (FS) shape and topology per valley--spin found in different regions of the parameter space. The Lifshitz transition produce distinct features in the DOS.
    }
    \label{fig:noninteracting_bands}
\end{figure*}

\textit{Bernal bilayer graphene band structure.} In \cref{fig:noninteracting_bands}(a-b), the electronic band structure of $h(\vec{k})$ in \cref{eq:continuum_hamiltonian_Nlayer} is strongly trigonally-warped (by interlayer hopping amplitudes) and has van Hove singularities near charge neutrality. The dispersion is semimetalic in absence of a displacement potential $U_D$ but has a insulating gap when it is finite. Typical experimentally accessible parameters are electronic density ranges of $n_e \in[-1\cdot 10^{12},0]\;\text{cm}^{-2}$, and displacement potentials in the range $U\in [0,80]$ meV. 
In \cref{fig:noninteracting_bands}(c), the topology and shape of the Fermi sea (FS) undergoes Lifshitz transitions when the electron density $n_e$ and the potentials $U_D$ changes over the experimentally accessible range: from a single simple FS to an annular FS, and then to three pockets. Each transition is accompanied by a van-Hove singularity that induces a discontinuity in the density of states, see \cref{fig:noninteracting_bands}(d): a jump in the case of going from single FS to annular FS, and a  saddle-point log-divergence when tuning towards isolated pockets.

\textit{Flavor magnetism.} Experimental and theory work on top/back-gated rhombohedral multilayers demonstrate that correlated states such as spin--valley magnetism, superconductivity and fractional anomalous quantum Hall states are ubiquitous in these materials \cite{Weitz2010,Velasco2012,Bao2011,Lee2014,Chen2019Mott,Chen2019SC,Chen2020Chern,Yang2022Spectroscopy,Shi2020RG,Zhou2021HalfQuarter,Zhou2021HalfQuarter,zhang_spinorbit_2022,Zhou2022Isospin,lu2023fractional,Han2024Correlated,Han2023Multiferroic,Han2024QAH,wolf2024magnetismdilute2deg,Zhang2010ABC,Koshino2009Trigonal,Huang2023Magnetism,Qin2023FRG,Cea2022KohnLuttinger,Dong2023Isospin,Ghazaryan2021Annular,Koh2024SOC,Lu2022RG,Dong2024SpinCanting}. The commonly-used effective intralayer Coulomb potential is $V_{\vec{q}}=(2\pi k_e/\epsilon_r) \tanh(\vert\vec{q}\vert d_{\text{gate}})/\vert\vec{q}\vert$, and accounts for dual gating with metallic gates at distance $d_{\text{gate}} \in [30,60]$~nm, and  $k_e=1.44$~eV~nm is the Coulomb constant. The phenomenological relative permittivity $\epsilon_r\in[10,35]$ is a tuning parameter in mean-field calculations, to avoid unrealistically large band distortions and band splittings. 
In \cref{fig:trilayer_graphene_fp}, we show the band structure of self-consistent Hartree-Fock mean-field states that compete as lowest-energy solutions for RTG.

\textit{Landau quantization.} Transverse magnetic fields quantize the electronic spectrum into Landau levels, described by a minimal coupling $\vec{p}\mapsto\vec{\pi}=\vec{p}-e\vec{A}$ in \cref{eq:continuum_hamiltonian_Nlayer}. The commutation relation of $\pi=\pi_x+i \pi_y$ with $\pi^\dagger$ implies ladder operators $\hat{a}$ and $\hat{a}^\dagger$, which we represent with matrices $\left(\hat{a}^{\dagger}\right)_{mn}=\bra{m}\hat{a}^{\dagger}\ket{n} = \delta_{m,n+1}\sqrt{n}$, 
with $\left[\hat{a},\hat{a}^\dagger\right]=1$ such that $\pi_x=\frac{\hbar}{\sqrt{2} l_B}\left(\hat{a}^{\dagger}+\hat{a}\right)$, $\pi_y=\frac{\hbar}{i\sqrt{2} l_B}\left(\hat{a}^{\dagger}-\hat{a}\right)$. We defined the magnetic length $\ell_B=\sqrt{\hbar/eB}$. Each Landau level has degeneracy $\Phi/\phi_0$, where $\Phi=B A$ is the flux and $\phi_0=h/e$ a flux quantum, and thus accommodates the electronic density $n_B=1/(2\pi \ell_B^2)$. Writing the Hamiltonian in the eigenbasis of $\hat{N}=\hat{a}^{\dagger}\hat{a}$ allows to obtain the electronic spectrum (numerically and in simplified cases analytically), see \cref{fig:bilayer_graphene_LL_suscept}(a) for BBG. We note for multilayer graphene some care is needed to avoid unphysical low-energy levels related to the Landau level truncation $N_{\text{c}}$ (in our case typically chosen between $50$ and $100$), which we deal with by removing any eigensolutions whose wavefunction has the majority of its weight at the basis truncation.

\subsection*{Additional multilayer graphene polarizability results at zero \texorpdfstring{$B$}{B}}

\begin{figure*}
  \centering
  \includegraphics[width=0.77\textwidth]{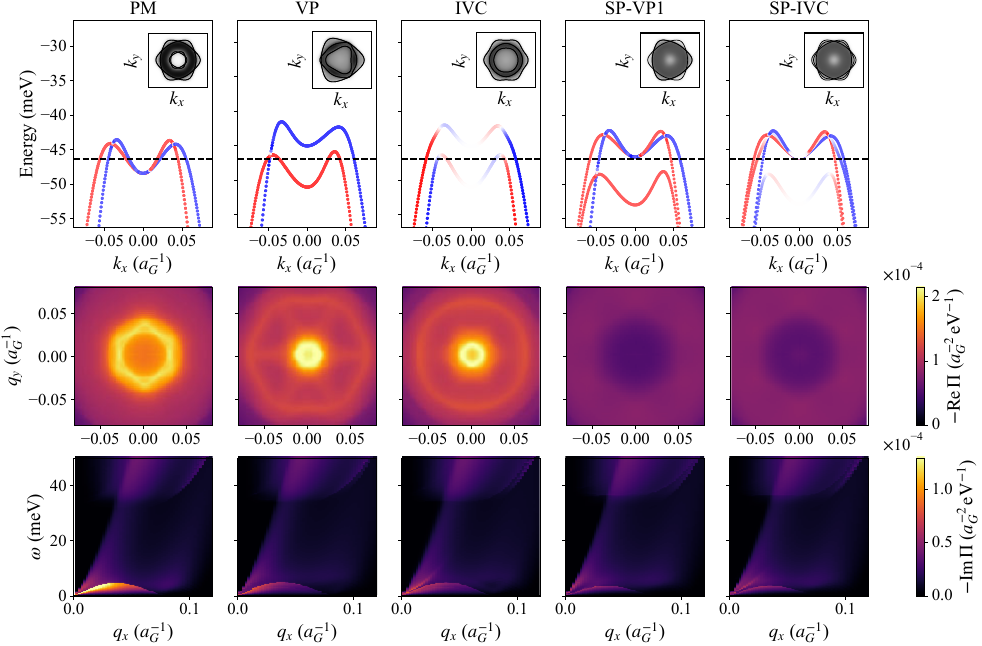}
  \caption{
  Overview of self-consistent mean-field band structures and corresponding single-particle susceptibility $-\chi_0$ (RPA polarizability $-\Pi$) for rhombohedral trilayer graphene at $n_e \!=\! -0.97\times10^{12}\,\mathrm{cm}^{-2}$ and $U_D \!=\! 30\,\mathrm{meV}$. Columns correspond to the different Hartree--Fock states (symmetric/PM, VP, IVC, SP-VP1, SP-IVC). Top row: mean-field band structures along $k_x$ with Fermi-surface insets. Middle row: static $-\mathrm{Re}\,\Pi(q_x,q_y,\omega\!\approx\!0)$. Bottom row: dynamical $-\mathrm{Im}\,\Pi(q_x,q_y\!=\!0,\omega)$. Momenta are in $a_G^{-1}$, $\omega$ in meV, and $\Pi$ in $a_G^{-2}\,\mathrm{eV}^{-1}$.
  Discretization: uniform $272\times272$ mesh, $|k_{x,y}|\!\le\!0.28\,a_G^{-1}$;  $n_\omega\!=\!1500$ uniform frequencies in $[0,150]$~meV; $1$~meV Lorentzian broadening.
  }
  \label{fig:trilayer_graphene_fp}
\end{figure*}

\begin{figure*}
  \centering
  \includegraphics[width=0.8\textwidth]{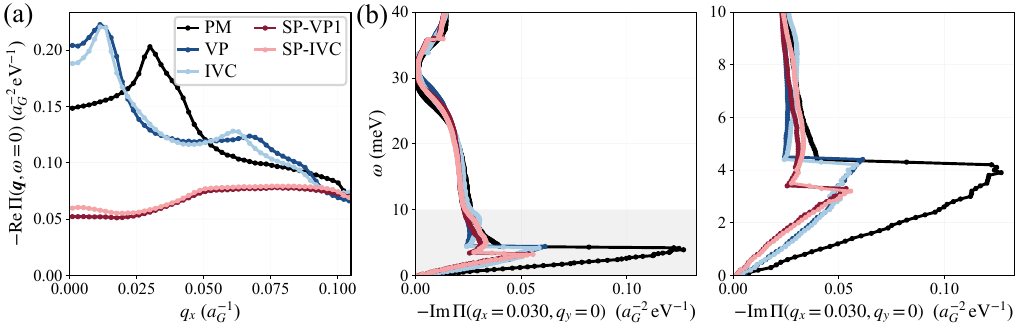}
  \caption{(a)~Static single-particle susceptibility (RPA polarizability $\Pi=\chi_0$) of rhombohedral trilayer graphene: line cuts of data shown in \cref{fig:trilayer_graphene_fp}. (b)~Dynamical non-interacting susceptibility $-\mathrm{Im}\,\chi_0(q_x = 0.03\,a_G^{-1}, q_y = 0, \omega)$ versus $\omega$; left: plot range ($0$--$40\,\mathrm{meV}$), right: zoom ($0$--$10\,\mathrm{meV}$). The gray band in (b, left) marks the zoom window. Momenta are in $a_G^{-1}$ and $\chi_0$ in $a_G^{-2}\,\mathrm{eV}^{-1}$. }
  \label{fig:rtg-chi0-three-panel}
\end{figure*}

\Cref{fig:trilayer_graphene_fp} summarises the mean-field electronic structure and the corresponding single-particle charge response of the \emph{symmetric/PM} and symmetry-broken Hartree–Fock states of rhombohedral trilayer graphene (RTG) at displacement field $U_D\!=\!30$\,meV and density $n_e\!=\!-0.97\times10^{12}\,\mathrm{cm}^{-2}$. 
The top-row panels reproduce the strongly trigonally-warped band dispersion and the annular Fermi surface that appears at this carrier density. 
Because of the near-nesting between the inner and outer Fermi contours, the static polarizability develops harp maxima at specific wavevectors, as seen in the middle-row panels showing $\mathrm{Re}\,\Pi(\vec q,0)$. 
The corresponding particle–hole continuum is visible in the bottom-row panels, where $\mathrm{Im}\,\Pi$ extends down to small frequencies because of available intraband transitions. 
These trends are quantified in \Cref{fig:rtg-chi0-three-panel}: panel~(a) displays line cuts of the static peaks in $\mathrm{Re}\,\Pi$ versus $q_x$, while panel~(b) shows that $\mathrm{Im}\,\Pi$ at fixed momentum carries substantial low-frequency spectral weight. 
When Coulomb interactions are included at the RPA level, these low-energy transitions are efficiently screened and evolve into a well-defined acoustic plasmon branch in the full RPA susceptibility $\chi$ (not shown), which underlies the sizable retardation corrections discussed below.

\Cref{tab:bandshift_rtg} collects the band-gap renormalisation
$\delta E_g$ induced in a proximate monolayer WSe$_2$ for each of the
competing Hartree–Fock solutions shown in
\cref{fig:trilayer_graphene_fp}.
The first two columns separate the static contribution
($\delta E_g^{\text{stat}}$) from the dynamic
retardation correction ($\delta E_g^{\text{dyn}}$) obtained from
\cref{eq:finalenergygap3,eq:finalenergygap_corr},
respectively.
The dynamical screening reduces the
magnitude of the gap shift compared to the static screening contribution. 
More importantly,
the \emph{difference} between flavour-polarised and paramagnetic states
originates from changes in both $\mathrm{Re}\,\chi$ and the low-frequency part of
$\mathrm{Im}\,\chi$: flavor polarisation suppresses the nested portions
of the Fermi surface, thereby lowering the static screening and leading
to a \emph{larger} (less negative) bandgap in the sensing layer.
We emphasise that these relative shifts, rather than the absolute
values, are the quantities that can be compared directly to experiment.

\begin{table}[h!]
\centering
\caption{WSe$_2$ bandgap change [cf.~\cref{eq:finalenergygap3,eq:energygap}] due to proximity to rhombohedral trilayer graphene in a strong displacement potential $U_D=30$~meV at density $n_e=-0.97\times 10^{12}$~cm$^{-2}$. Here, we consider the case without a spacer layer, $d\approx 2.4 a_G$.  The Hartree-Fock state labels refer to \cref{fig:trilayer_graphene_fp}.}
\label{tab:bandshift_rtg}
\begin{tblr}{
  width = 0.6\linewidth,
  colspec = {X *{4}{X}},
  rowsep = 1pt,
  column{1} = {leftsep=0pt},
  column{Z} = {rightsep=0pt},
  hline{1,Z} = {1pt,solid},
  hline{2,4,6} = solid,
}
 HF state & $\delta E_{\text{g}}^{\text{stat}}$ & $\delta E_{\text{g}}^{\text{dyn}}$ & $\delta E_{\text{g}}$ & $\delta E_{\text{g}}-\delta E_{\text{g}}^{\mathrm{PM}}$ \\
 PM     & $-14.27$ &  $8.50$ & $-5.77$ & $0.00$ \\
 VP     & $-14.25$ &  $9.73$ & $-4.52$ & $1.25$ \\
 IVC    & $-14.25$ &  $9.70$ & $-4.55$ & $1.22$ \\
 SP-VP1 & $-13.94$ &  $9.58$ & $-4.36$ & $1.42$ \\
 SP-IVC & $-13.97$ &  $9.53$ & $-4.43$ & $1.34$ \\
\end{tblr}
\end{table}

\subsection*{Additional multilayer graphene polarizability results at finite \texorpdfstring{$B$}{B}}

\begin{figure*}
  \centering
  \includegraphics[width=0.9\linewidth]{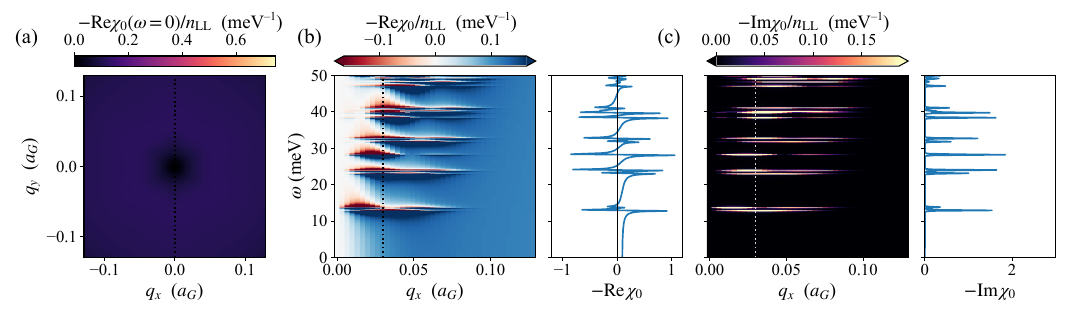}
  \caption{%
    Non–interacting susceptibility $\chi_{0}$ of spin–valley–resolved Bernal bilayer graphene at $B=3$\,T when the $n=0$ \emph{and} $n=1$ Landau-level orbitals are empty (\emph{even} filling).  
    (a)~Static contribution $-\mathrm{Re}\,\chi_{0}(\omega\!=\!0)/n_{\text{LL}}$ in the $(q_{x},q_{y})$-plane shows a single, featureless peak centred at $\vec q=\vec 0$, characteristic of weak screening.  
    (b)~Frequency–resolved $-\mathrm{Re}\,\chi_{0}/n_{\text{LL}}$ along $q_{y}\!=\!0$ reveals inter-LL particle–hole continua with onsets at $\omega\gtrsim 2\Delta_{01}$; the marginal panel (right) displays the cut at $q_{x}=0.03\,a_{G}^{-1}$.  
    (c)~Imaginary part $-\mathrm{Im}\,\chi_{0}/n_{\text{LL}}$ for the same momentum cut, highlighting the absence of low-energy spectral weight.%
  }
  \label{fig:bilayer_graphene_chi0_filling0}
\end{figure*}

\begin{figure*}
  \centering
  \includegraphics[width=0.9\linewidth]{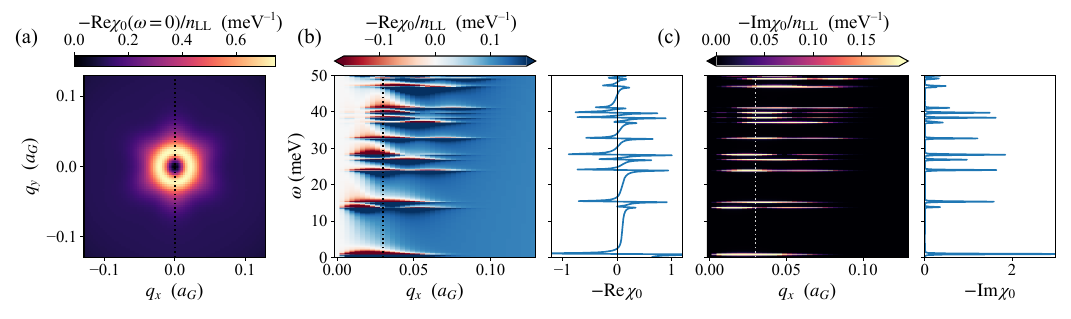}
  \caption{%
    Same as Fig.~\ref{fig:bilayer_graphene_chi0_filling0}, but for \emph{odd} filling where the $n=0$ Landau level is fully occupied while $n=1$ remains empty.  
    (a)~In the static response a pronounced \emph{annular} maximum develops at $q\!\sim\!\ell_{B}^{-1}$, signalling enhanced screening due to $n=0\!\rightarrow\!1$ excitations.  
    (b)~$-\mathrm{Re}\,\chi_{0}$ now exhibits a broad ridge at $\omega\!\approx\!\Delta_{01}$ that disperses weakly with $q_{x}$; the associated peak is clearly visible in the marginal cut.  
    (c)~The corresponding ridge in $-\mathrm{Im}\,\chi_{0}$ reflects low-energy particle–hole states which, after RPA resummation, drive the filling-factor dependence of the WSe$_2$ band-gap shift.
  }
  \label{fig:bilayer_graphene_chi0_filling1}
\end{figure*}

\Cref{fig:bilayer_graphene_chi0_filling0,fig:bilayer_graphene_chi0_filling1} present the non-interacting
susceptibility $\chi_{0}(\vec q,\omega)$ that is fed into the RPA
screening calculation discussed in the main text.
Both figures employ the same Bernal bilayer-graphene parameters used in
\cref{fig:bilayer_graphene_LL_suscept} but compare two distinct
occupations of the eightfold Landau-level (LL) octet at $B=3$\,T:

\begin{itemize}
  \item \emph{Even filling} — the $n=0$ \emph{and} $n=1$ orbitals are
        empty (chemical potential in the inter-octet gap),
        \cref{fig:bilayer_graphene_chi0_filling0}.
  \item \emph{Odd filling}  — the $n=0$ orbital is completely filled
        while $n=1$ remains empty (chemical potential inside the octet),
        \cref{fig:bilayer_graphene_chi0_filling1}.
\end{itemize}

For the even case the static response
$-\mathrm{Re}\,\chi_{0}(\vec q,0)$ is nearly featureless and peaks only
at $\vec q=\vec 0$, indicating weak screening.
Correspondingly, the dynamic response shows particle-hole excitations with
onsets at $\omega\gtrsim2\Delta_{01}$ and practically no spectral weight
at low frequencies. 

Once the $n=0$ LL is occupied the situation changes qualitatively:
$-\mathrm{Re}\,\chi_{0}(\vec q,0)$ develops an annular maximum at
$q\!\sim\!\ell_{B}^{-1}$, and a pronounced ridge appears in
$-\mathrm{Re}\,\chi_{0}$ (and likewise in
$-\mathrm{Im}\,\chi_{0}$) at
$\omega\!\approx\!\Delta_{01}$.
These new low-energy particle-hole excitations are responsible for the
sharp increase of the static polarizability and, after RPA resummation,
drive the strong filling-factor dependence of the WSe\textsubscript{2}
band-gap shift reported in \cref{fig:bilayer_graphene_LL_suscept}(c,d) of the
main text.
The comparison therefore emphasises that \emph{even subtle changes in
Landau-level filling leave clear fingerprints in the proximate
susceptibility and can be detected optically via the exciton sensor
layer.}

\end{document}